\documentclass[useAMS,usenatbib]{article}
\usepackage{hyperref}
\usepackage{graphicx}

\begin{document}

\title{A Python Code for the Emmanoulopoulos et al. 2013 Light Curve Simulation Algorithm}

\author{ S. D. Connolly\footnote{contact: sdc1g08@soton.ac.uk} \\ \\
\small{Physics and Astronomy, University of Southampton, SO17 1BJ}\\
}

\date{}

\maketitle

\begin{abstract}

\noindent I have created a python code for the light curve simulation algorithm of Emmanoulopoulos et al. 2013, which is available at:\\

\url{https://github.com/samconnolly/DELightcurveSimulation}\\

\end{abstract}

\section{Description}
\begin{figure*}
	\includegraphics[width = \columnwidth]{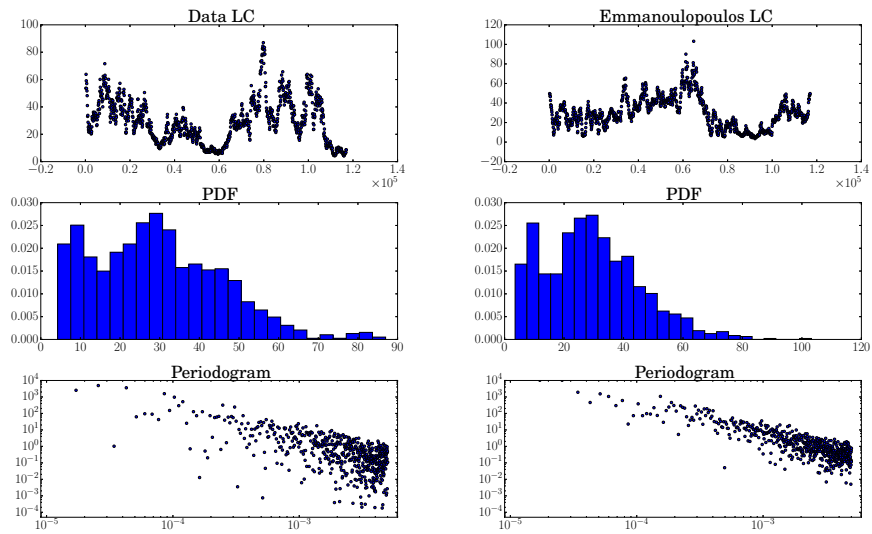}
	
	\caption{Lightcurves, PDFs and periodograms of (left) a real data set and (right) an artificial light curve produced using the Emmanoulopoulos et al. 2013 method.
	Both the PDF and the periodogram of the simulated light curve are seen to be the same as those of the original data set.}
	
	\label{lcs}
\end{figure*}

\noindent I have created, for public use, a Python code allowing the simulation of light curves with
any given power spectral density and any probability density function (PDF), following the algorithm described
in Emmanoulopoulos et al. 2013.
The simulated products have exactly the same variability and statistical properties as the observed light curves (see Fig. \ref{lcs}).

Note that a Mathematica code of the algorithm is given in Emmanoulopoulos et al. 2013.

\end{document}